\documentclass[a4paper]{article}
\usepackage{epsfig}
\usepackage[dvips]{color}

\begin{document}
\title{Enhancement of the escape time in metastable states with
colored noise}
\author{A.Fiasconaro\footnote{E-mail adress: afiasconaro@gip.dft.unipa.it}, D.Valenti, B.Spagnolo\\
{\em \normalsize INFM and Dipartimento di Fisica e Tecnologie
Relative} \\ {\em \normalsize Viale delle Scienze - 90128 Palermo,
Italy}}
\date{today}
\maketitle
\begin{abstract}
We present a study of the escape time from a metastable state in
the presence of colored noise, generated by Ornstein-Uhlenbeck
process. We analyze the role of the correlated noise and of
unstable initial conditions of an overdamped Brownian particle on
the enhancement of the average escape time as a function of the
noise intensity. We observe the noise enhanced stability (NES)
effect for all the initial unstable states and for all values of
the correlation time $\tau_c$ investigated. We can distinguish two
dynamical regimes characterized by: (a) a weak correlated noise
and (b) a strong correlated noise, depending on the value of
$\tau_c$ with respect to the relaxation time. With increasing
$\tau_c$ we find : (i) a shift of the maximum of the average
escape time towards higher values of noise intensity and an
enhancement of the value of this maximum; (ii) a broadening of the
NES region, which becomes very large in the strong colored noise
regime; (iii) in this regime (b), the absence of the peculiar
initial condition $x_c$ which separates the set of the initial
unstable states lying into the divergency region from those which
give only a nonmonotonic behavior of the average escape time.
\end{abstract}
%
%\begin{keyword}
%Statistical Mechanics, Mean First Passage Time, Noise-induced effects, NES
%\PACS 05.40-a,87.23Cc,89.75-k
%\end{keyword}
%

\section{Introduction}
Nonlinear relaxation decay of physical systems from an initial
unstable or metastable state involves fundamental aspects of
non-equilibrium statistical mechanics. The investigation of
nonlinear dynamics and instabilities in systems away from
equilibrium in fact has led to some counterintuitive and
resonance-like phenomena. Among these we cite the stochastic
resonance \cite{Gam}, the resonant activation \cite{Doe}, the
noise enhanced stability (NES) \cite{Man,Agu} and noise induced
phase transitions \cite{Fre}. Recent theoeretical investigations
have shown that the average escape time from metastable states in
fluctuating potentials has a nonmonotonic behaviour as a function
of the noise intensity \cite{Agu,Ale}. This resonance-like
behaviour, which contradicts the monotonic behaviour predicted by
the Kramers theory \cite{Kra,Han,Hir}, is the NES phenomenon: the
stability of metastable or unstable states can be enhanced by the
noise and the average life time of the metastable state is greater
than the deterministic decay time.

The inclusion of realistic noise sources with a finite correlation
time in modelling dynamical systems can impact both the stationary
and the dynamic features of nonlinear systems. For metastable
thermal equilibrium systems it has been demonstrated that colored
thermal noise can substantially modify the diffusive barrier
transmission \cite{Han}. For bistable systems the colored noise
driven escape rate for small and large correlation time has been
derived in ref. \cite{Han1}. A rich and enormous literature on
escape processes driven by colored noise was produced in the 80's.
As an example of this literature concerning colored noise we cite
the escape from metastable states \cite{new}, the MFPT from a
marginal state \cite{Ramirez} and the decay of an unstable state
\cite{Sancho}. In this work we present a study of the average
decay time of an overdamped Brownian particle subject to a cubic
potential with a metastable state. We focus on the role of
different unstable initial conditions and of colored noise in the
average escape time. The effect of \emph{color} on the transient
dynamics of the escape process is strictly related to the
characteristic time scale of the system, i. e. the relaxation time
inside the metastable state $\tau_r$. For correlation time values
less than the relaxation time ($\tau_c < \tau_r$), the system
"sees" the noise source as white noise and the dynamical regime of
the Brownian particle is like the white noise dynamics with a
shift of all curves calculated at different initial positions
towards higher values of noise intensity. For strong color, i. e.
for values of correlation time larger than the relaxation time
($\tau_c > \tau_r$) nontrivial results are obtained. Particularly
we obtain: (i) a big shift of the behaviours of the average escape
times towards higher noise intensities; (ii) an enhancement of the
value of the average escape time maximum and a broadening of the
NES region in the plane ($\tau, D$), which becomes very large for
high values of $\tau_c$; (iii) the absence of the peculiar initial
position $x_c$, found in a previous study \cite{Ale}, which
separates the set of the initial unstable states producing
divergency from those which give only a nonmonotonic behavior of
the average escape time. This is the relevant result of this
paper. For all the initial unstable states we find an enhancement
of the average escape time with respect to the deterministic decay
time and in the strong correlated noise regime ($\tau_c >>
\tau_r$) a nonmonotonic behavior of the the average escape time.
\section{The model}
The starting point of our study is the Langevin
 equation

 \begin{equation}
  \dot{x}=-\frac{\partial U(x)}{\partial x} + \eta(t)
  \label{eq}
 \end{equation}
where $\eta(t)$ is the Ornstein-Uhlenbeck process \cite{Orn,Hor}

 \begin{equation}
  d\eta=-k \eta dt + k \sqrt{D} \; d W(t)
  \label{eqou}
 \end{equation}
where $W(t)$ is the Wigner process giving a Gaussian noise with
the usual statistical properties:
 $\langle \xi(t) \rangle = 0$ and $\langle \xi(t)\xi(t+\tau)
 \rangle = \delta
 (\tau)$.
The integration of Eq.(\ref{eqou}) yields

 \begin{equation}
  \eta(t)=\eta(0)e^{-kt}+k\sqrt{D} \int_0^t e^{-k(t-t')} dW(t')
 \label{eqetat}
 \end{equation}
which for $\eta(0) = 0$ and for $t \to \infty$ gives

 \begin{equation}
  \langle \eta(t) \rangle =  0,
   \label{meanou}
 \end{equation}

 \begin{equation}
   \langle \eta(t) \eta(t+\tau) \rangle = \frac{k D}{2}
   e^{-k\tau},
    \label{corou}
 \end{equation}
where $1/k=\tau_c$ is the correlation time of the process.
 The system of stochastic differential equations (\ref{eq}) and
 (\ref{eqou}), which represent a two-dimensional Markovian process,
 is equivalent to a non-Markovian Langevin equation
 driven with additive Gaussian correlated noise with $\eta(t)$
 obeying the properties (\ref{meanou}) and (\ref{corou}).

\begin{figure}[htbp]
 \begin{center}
  \includegraphics[height=7cm]{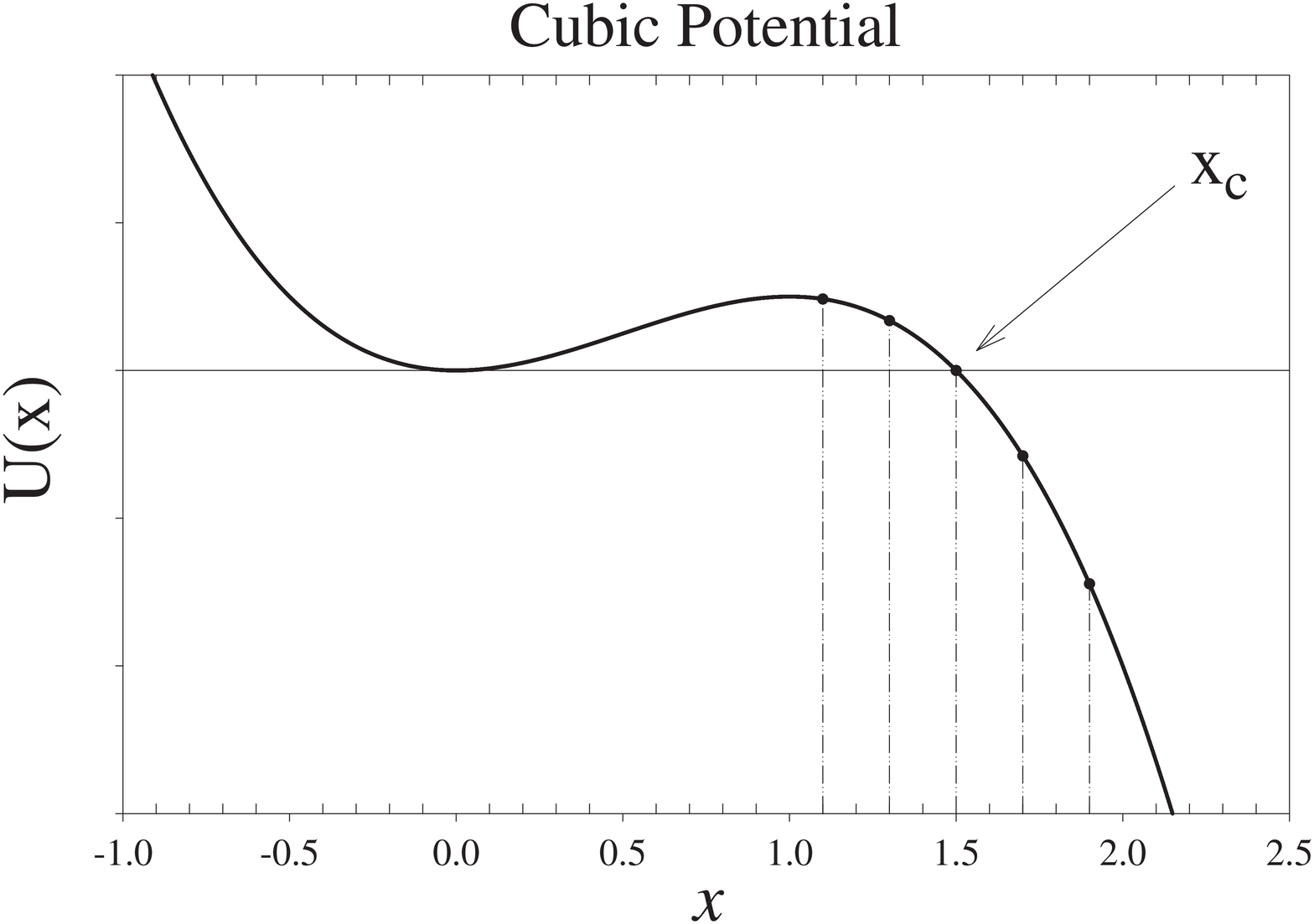}
  \linespread{1.2} %interlinea
  \caption{The cubic potential $U(x)$  the various initial
  positions investigated (dots); $x_c$ is the critical initial position
  found in the case of white noise, which remains only for weak colored
  noise ($\tau_c < \tau_r$). The absorbing boundary is $x_F = 20$.}
  \label{fpot}
 \end{center}%
\end{figure}

It is possible to show that the Eq.(\ref{eqou}) for the OU process
gives the correct limit $\lim_{\tau_c \to 0}
 \eta(t) = \sqrt{D} \xi(t)$, i. e. the white noise term.
In fact

 \begin{eqnarray}
  \lim_{\tau_c \to 0} \eta(t) & = & 2\sqrt{D} \int_0^t \lim_{\tau_c \to 0} \frac{e^{-(t-t')/\tau_c}}{2\tau_c}
  dW(t')
  \nonumber \\  &=& 2\sqrt{D} \int_0^t \delta(t-t') \xi(t')d(t')
   = \sqrt{D} \xi(t),
 \label{etlim}
  \end{eqnarray}
where we use the property of $\delta$ function

\begin{equation}
  \int_{-\infty}^{+\infty} dx \delta(x - x_o) f(x) = f(x_o),
   \label{deltaf}
 \end{equation}
 and the factor $2$ disappears because of the integration range. The
stationary correlation function of the Ornstein-Uhlenbeck process
(\ref{corou}) gives in the limit $\tau_c \to 0$ the correlation
function of the white noise

 \begin{equation}
  \lim_{\tau_c \to 0} \langle \eta(t) \eta(t+\tau) \rangle =
  \lim_{\tau_c \to 0} \frac{D}{2\tau_c}
  e^{-\tau/\tau_c}= D\delta(\tau).
  \label{corou1}
 \end{equation}
 The potential $U(x)$ used in Eq.(\ref{eq}) is
a cubic one

 \begin{equation}
  U(x)=ax^2 - bx^3,
  \label{u1}
 \end{equation}
 with $a=0.3$, $b=0.2$.
The potential profile has a local stable state at $x_0=0$  and an
unstable state at $x_0=1$ (see Fig.\ref{fpot}).

The relaxation time for the metastable state at $x_0=0$ is

\begin{equation}
  \tau_r = \left[\frac{d^2 U(x)}{dx^2}\right]_{x=0} = 2 a,
  \label{u1}
 \end{equation}
which is the characteristic time scale of our system.

\section{Results and Comments}
\begin{figure}[htbp]
 \centering
 \includegraphics[height=7.5cm]{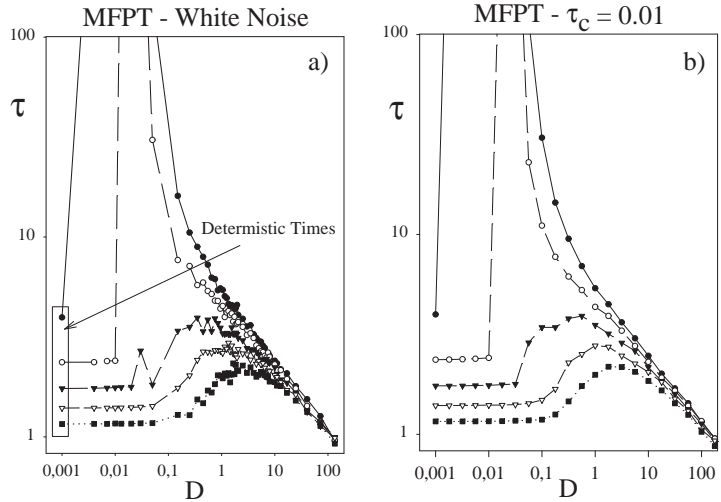}
% \linespread{0.7} %interlinea
 \vskip -0.2cm
 \caption{ a) Log-Log plot of the Mean First Passage Time as a function of noise
intensity in the case of white noise for the five initial
positions investigated in this paper (see Fig.\ref{fpot}), namely:
$x_o=1.1 \div 1.9$, with steps of $0.2$ \cite{Ale}; b) MFPT for
$\tau_c = 0.01$ and the same initial positions.}
 \label{fnes}
\end{figure}

\begin{figure}[htbp]
 \centering
 \includegraphics[height=13cm]{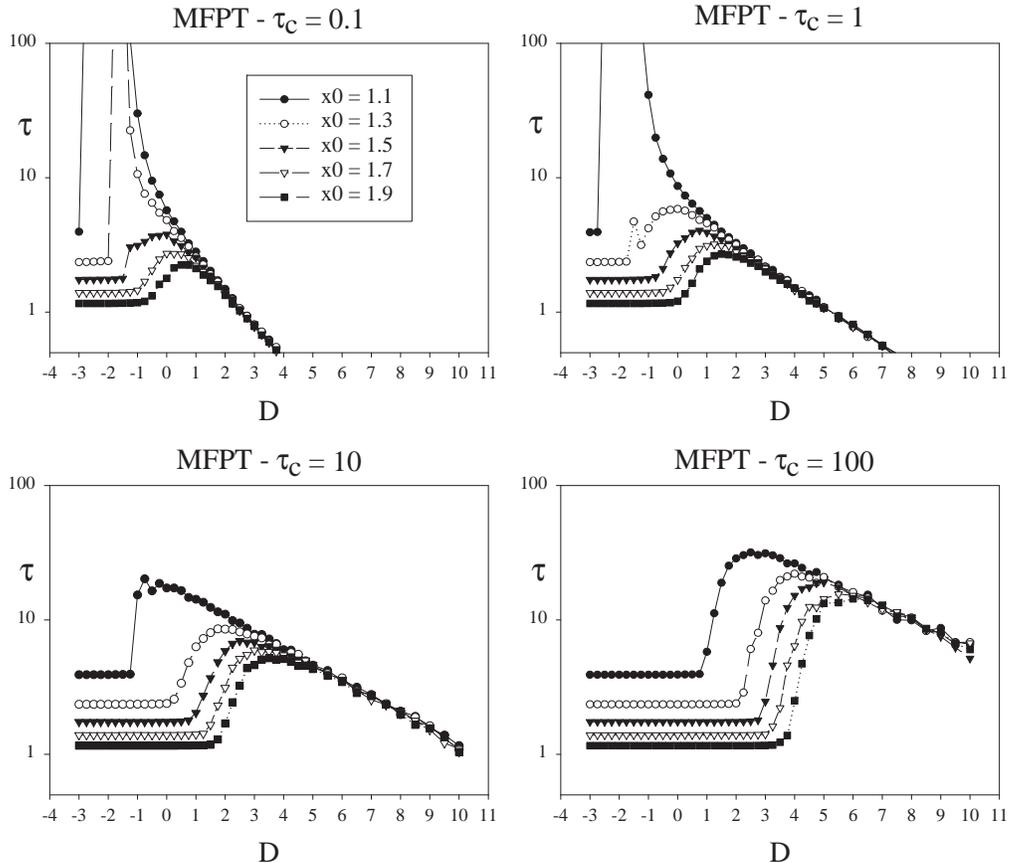}
 \linespread{1.2} %interlinea
 \vskip -0.2cm
 \caption{Log-Log plot of the Mean First Passage Time as a function of noise
intensity for the five initial positions investigated (see
Fig.\ref{fpot}) and for four values of the correlation times
$\tau_c$. In the $x$-axes we report only the order of magnitude in
powers of ten.}
 \label{tau}
\end{figure}

The calculations of the average escape time as a function of the
colored noise intensity have been performed by averaging over
$5000$ realizations the stochastic differential equation
(\ref{eq}). The absorbing boundary for the escape process is put
on $x_F = 20$. For all the initial unstable states (see
Fig.\ref{fpot}) and all the correlation times evaluated we find an
enhancement of the average escape time with respect to the
deterministic time as a function of the noise intensity. We can
observe the characteristic behavior of the NES effect for unstable
states with white noise in Fig.\ref{fnes}a \cite{Ale} where we
also find the comparison with the calculation performed with
colored noise with $\tau_c=0.01$ (Fig.\ref{fnes}b). The behaviours
of the average escape time as a function of colored noise
intensity with the others values of $\tau_c$ are shown in
Fig.\ref{tau}.

We clearly observe two dynamical regimes: (a) weak colored noise
($0<\tau_c<\tau_r$) and (b) strong colored noise ($\tau_c >
\tau_r$). Starting our observation from Fig.\ref{tau} we see that
the qualitative behavior of MFPT with white noise is recovered
with $\tau_c = 0.1$, until $\tau_c \simeq \tau_r = 0.6$. That is
an enhancement of the average escape time with respect to the
deterministic decay time with a divergent behaviour for
$x_{max}<x_0<x_c$ and a non monotonic behavior for $x_0 \ge x_c$.
Increasing the value of the correlation time Fig.\ref{tau}
($\tau_c \ge \tau_r$) we observe a shift of the maximum towards
higher values of the noise intensity and an enhancement of the
value of the average escape maximum. Moreover we observe a
broadening of the NES region, which becomes very large (up to
twelve order of magnitude) for high values of the correlation time
$\tau_c$ (see Fig.3 with $\tau_c = 100$). The unexpected result is
that increasing the correlation time the divergent behaviour tends
to disappear, as shown in Fig.\ref{tau} for $\tau_c=1.0$ and
initial position $x_o=1.73$, and disappear for all the initial
positions investigated for $\tau_c=10$ and $\tau_c=100$.
 For high values of the noise intensity all the graphs
show a monotonic decrease behavior as a function of noise
intensity collapsing in a unique curve, according to the behavior
predicted by the Kramers theory \cite{Kra,Han}. Moreover
 the slope of this limit curve
 becomes  flatter increasing the
correlation time. This means that the NES effect involves more and
more orders of magnitude of the noise intensity. The effect of the
colored noise is therefore to delay the escape process or in other
words to enhance more and more the stability of the metastable
state.

\section{Conclusions}

 In this work we analyzed the effect of the colored noise,
generated by an OU process, on the enhancement of the average
escape time in a cubic potential with a metatstable state. We
analyze different initial unstable states with $x_0 > x_c$, where
$x_c$ is the crossover between the divergent behaviour and the
nonmonotonic behaviour of the average escape time \cite{Agu,Ale}.
We obtain NES effect for all the initial positions investigated
and an enhancement of the NES effect for increasing values of
correlation times. The results obtained for a particle moving in a
cubic potential are quite general, because we always obtain NES
effect when a particle is initially located just to the right of a
local potential maximum, with a local minimum or metastable state
in its left side and the global escape region in its right side.

 From a comparison with analogue calculation with
white noise, we observe the following similarity and differences
for increasing values of the correlation times: 1) the peculiar
point $x_c$ found for white noise separating divergent behaviours
from non-monotonic behaviour, is present only in the weak colored
regime $\tau_c < \tau_r$; 2) all the divergencies present in the
case of white noise for all the unstable initial position in the
range $x_{max}<x_0<x_c$ tend to transform in pure NES effect, i.
e. without divergencies, for increasing values of $\tau_c$; 3) the
region in the plane $(\tau,D)$ of the NES effect increases up to
$12$ order of magnitude for increasing values of $\tau_c$ (strong
colored noise regime).

Nonmonotonic behavior of the mean escape time as a function of
noise intensity therefore is a noise-induced effect for nonlinear
nonequilibrium systems with metastable states which is enhanced
when we consider realistic noise sources with finite correlation
time.

In experiments real noise sources are correlated with a finite
correlation time. As a consequence the NES effect can be observed
at higher noise intensities with respect to the idealized white
noise case. Therefore the theoretical results obtained using white
noise source are only a good approximation for low values of the
correlation time. The enhancement and the shift of the NES region
towards higher values of the noise intensity allow to reveal
experimentally the NES effect using a suitable correlation time
$\tau_c$.

\section{Acknowledgments}
 This work was
supported by \mbox{INTAS Grant 01-0450}, by INFM and MIUR.

\end{document}